\documentclass{amsart}
\usepackage{graphicx,epsfig}
\usepackage{multirow}
\usepackage{algorithm}
\usepackage{algpseudocode}
\usepackage{mathrsfs}
\usepackage{amsmath,amssymb,latexsym, amsfonts, amscd, amsthm, xy}
\input{xy}
\xyoption{all}
\usepackage{natbib}
\input{xy}
\xyoption{all}

\numberwithin{equation}{section}

\usepackage[usenames]{color}

\usepackage{parskip}

\makeindex
=651pt \headheight = 12pt \headsep = 0pt

\theoremstyle{plain}
\newtheorem{theorem}{Theorem}[section]

\theoremstyle{definition}

\newtheorem{remark}{Remark}[section]

\def\cM{\mathcal{M}}

\def\etr{\text{etr}}

\newcommand{\dif}{\mathrm{d}}

\def\ml{\text{P}_{\text{ML}}}

\def\etr{\text{etr}}
\DeclareMathOperator{\Tr}{Trace}
\DeclareMathOperator{\diag}{Diag}
\def\bkappa{\boldsymbol\kappa}

\def\cM{\mathcal{M}}

\def\etr{\text{etr}}

\title{Bayesian nonparametric inference on the Stiefel manifold}
\author{Lizhen Lin \and Vinayak Rao \and David B. Dunson}

\address{Department of Statistical Science, Duke University, USA }

\email{lizhen@stat.duke.edu, var11@stat.duke.edu, dunson@duke.edu}

\thanks{This work was supported by grant R01ES017240 from the National Institute of Environmental Health Sciences (NIEHS) of the National Institute of Health (NIH).}

\begin{document}
\maketitle

\begin{abstract}

The Stiefel manifold $V_{p,d}$ is the space of all $d \times p$ orthonormal matrices, with the $d-1$ hypersphere and the space of all orthogonal matrices constituting 
special cases.
In modeling data lying on the Stiefel manifold, parametric distributions such as the  matrix Langevin distribution are often used; however, model misspecification is a concern and it is 
desirable to have nonparametric alternatives.  Current nonparametric methods are Fr\'echet mean based.  We take a fully generative nonparametric approach, which relies on mixing parametric 
kernels such as the matrix Langevin.  The proposed kernel mixtures can approximate a large class of distributions on the Stiefel manifold, and we develop theory showing posterior 
consistency. While there exists work developing general posterior consistency results, extending these results to this particular manifold requires  substantial new theory. Posterior 
inference  is illustrated on a real-world dataset of near-Earth
objects.


\end{abstract}

\section{Introduction}
\label{intro}

Statistical analysis of matrices with orthonormal columns has diverse applications including
principal components analysis,  estimation of rotation matrices, as well as in analyzing orbit data of the orientation of comets and asteroids.
 Central to probabilistic models involving such matrices
are probability distributions on the Stiefel manifold,  the space of all $d \times p$ orthonormal matrices.
Popular examples of 
parametric distributions are the matrix  von Mises-Fisher distribution \citep{khatri1977, hornik2013}
(also known as the matrix Langevin \citep{chikuse1993,chikuse2003,chikuse2006}), and its
generalization, the Bingham-von Mises-Fisher distribution \citep{hoff2009}. Maximum likelihood estimation is often used in estimating the parameters, while recently 
\cite{rao14}  proposed a sampling algorithm allowing Bayesian inference for such distributions.
 
 Current parametric models are overly simple for most applications, and nonparametric inference has been limited to estimation of Fr\'echet means \citep{rabibook}.  
Model-based nonparametric inference has several advantages, including providing a fully generative model for prediction and characterization of uncertainty, while allowing adaptation to the 
complexity of the data. We propose a class of nonparametric models based on mixing parametric kernels on the Stiefel manifold.  Such models have appealing properties including large 
support, posterior consistency, and straightforward computation adapting the sampler of \cite{rao14}. Depending on the application, our models can be used to characterize the data directly, or to describe latent components of a hierarchical model.
 
%

  Section \ref{sec:geom}  provides some details on the geometry of the Stiefel manifold. Section \ref{sec:Bays_inf} introduces the matrix Langevin distribution, the nonparametric model and the posterior consistency theory. Section \ref{sec:post_sim} illustrates the model through application to an object orbits data set.  All proofs are included in the appendix.
  

\section{Geometry of the Stiefel manifold}
\label{sec:geom}

The Stiefel manifold $V_{p,d}$ is the space of all $p$-frames in $\mathbb{R}^d$, with a $p$-frame consisting of $p$ ordered orthonormal vectors in $\mathbb{R}^d$.
Writing $M(d,p)$ for the space of all $d \times p$ real matrices, and letting $I_p$ represent the $p \times p$  identity matrix, the Stiefel manifold can be
represented as
\begin{equation}
V_{p,d}=\{X\in M(d,p): X^TX=I_p\}.
\end{equation}
The Stiefel manifold $V_{p, d}$ has the $d-1$ hypersphere $S^{d-1}$ as a special case when $p=1$. When $p=d$, this is the space of all the orthogonal matrices
$O(d)$.
$V_{p,d}$  is a Riemannian manifold of dimension $dp-p-p(p-1)/2=p(2d-p-1)/2$. It can be embedded  into the Euclidean space $M(d,p)$ of dimension $dp$ with  the inclusion map as a natural embedding, and is thus a submanifold of $\mathbb R^{dp}$.

Let $G\in V_{p,d}$, and  $G_1$ be a matrix of size $d\times (d-p)$ such that $[G: G_1]$ is in $O(d)$, the group of $d$ by $d$ orthogonal matrices.
The volume form on the manifold is $\lambda(\dif G)=\wedge_{i=1}^p\wedge_{j=i+1}^{d}g_j^T\dif g_i$ where $g_1,\ldots,g_p$ are the columns of
$G$, $g_{p+1},\ldots, g_d$ are the columns of $G_1$ and $\wedge$ represents the wedge product \citep{muir2005}.  If $p=d$, that is when $G\in O(d)$,
one can represent $\lambda(\dif G)=\wedge_{i<j}g_j^T\dif g_i$. Note that $\lambda(\dif G)$ is invariant under the left action of the orthogonal group $O(d)$ and
the right action of the orthogonal group $O(p)$, and forms the Haar measure on the Stiefel manifold.
For more details on the Riemannian structure of the Stiefel
manifold, we refer to \cite{Edelman98thegeometry}.

\section{Bayesian nonparametric model }
\label{sec:Bays_inf}

Let $X$ be a random variable on $V_{p,d}$.  A popular parametric distribution of $X$ is  the matrix Langevin distribution which has the following  density $\ml$ with respect to the invariant Haar
volume measure on $V_{p,d}$
\begin{equation}
\label{eq-paraML}
\ml(X|F)=\etr(F^TX)/Z(F),
\end{equation}
The parameter $F$ is a $d\times p$ matrix, and the normalization constant $Z(F)=\mathstrut_0F_1(\frac{1}{2}d, \frac{1}{4}F^TF)$  is the hypergeometric
function  with matrix arguments, evaluated at $\frac{1}{4}F^TF$ \citep{chikusebook}.
Write the singular value decomposition (SVD) of $F$ as $F = G \bkappa H^T$, with $G$ and $H$, $d \times p$ and $p \times p$ orthonormal matrices, and $\bkappa$ 
a diagonal matrix with positive elements. One can think of $G$ and $H$ as orientations, with $\bkappa$ controlling the
concentration in the directions determined by these orientations.
Large values of $\bkappa$ imply concentration along the associated directions, while setting $\bkappa$ to zero recovers the uniform distribution on
the Stiefel manifold. \cite{khatri1977} show that $\mathstrut_0F_1(\frac{1}{2}d, \frac{1}{4}F^TF) =  \mathstrut_0F_1(\frac{1}{2}d, \frac{1}{4}\bkappa^T\bkappa)$, so that
the normalization constant depends only on $\bkappa$, and we write it as $Z(\bkappa$).
The mode of the distribution is  given by $GH^T$, and
from the characteristic function of $X$, one can show $E(X)=FU$, where the $(i,j)$th element of the matrix $U$ is given by
\begin{equation*}
U_{ij}=2\dfrac{\partial \log \mathstrut_0F_1(\frac{1}{2}d, \frac{1}{4}F^TF) }{\partial (F^{T}F)_{ij}}.
\end{equation*}

Consider $n$ observations  $X_1,\ldots,X_n$ drawn i.i.d.\ from $\ml(X|F)$. A simple approach to characterizing these observations is via a maximum likelihood
estimate of the parameter $F$ \cite[Section 5.2]{chikusebook}. Bayesian estimation of $F$ is on the other hand very challenging due to the intractable normalizing constant in the
likelihood. \cite{rao14}  proposes a sampling scheme based on  a data augmentation technique to solve this intractability problem.

In many situations, assuming the observations come from a particular parametric family such as matrix Langevin is restrictive, and raises concerns about model misspecification. Nonparametric alternatives, on the other hand, are  more flexible and have much wider applicability, and we consider these in the following. 

Denote by $\mathcal{M}$ the space all the densities on $V_{p,d}$ with respect to the Haar measure $\lambda$.  Let $g(X,G,\boldsymbol\kappa)$ be a parametric kernel on the Stiefel manifold
with a `location parameter' $G$ and a vector of concentration parameters $\boldsymbol\kappa=\{\kappa_1,\ldots, \kappa_p\}$.   One can place a prior $\Pi$ on $\mathcal{M}$ by modelling the random density $f$ as
\begin{equation}
\label{eq-mixmodel1}
f(X)=\int  g(X,  G, \boldsymbol\kappa) P(\dif \bkappa \dif G),
\end{equation}
with the mixing measure $P$ a random probability measure. A popular prior over $P$ is the Dirichlet process \citep{Fer1973}, 
parametrized by a base probability measure $P_0$ on the product space $\mathbb R_{+}^p\times V_{p,d}$,  and a concentration parameter $\alpha>0$.
We denote by $\Pi_1$ the DP prior on the space of mixing measures,
and assume $P_0$ has full support on $\mathbb{R}_{+}^p\times V_{p,d}$.

The  model in \eqref{eq-mixmodel1} is a `location-scale' mixture model, and corresponds to an infinite mixture model where each component has its
own location and scale. One can also define the following `location' mixture model given by
\begin{equation}
\label{eq-mixmodel2}
f(X)=\int  g(X,  G, \boldsymbol\kappa) P(\dif G)\mu(\dif \boldsymbol \kappa),
\end{equation}
where $P$ is given a nonparametric prior like the DP and $\mu(\dif \boldsymbol \kappa)$ is a parametric distribution 
(like the Gamma or Weibull distribution). In this model, all components are constrained to have the same scale parameters $\bkappa$.

When $\Pi_1$ corresponds to a DP prior, one can precisely quantify the mean of the induced density $\Pi$.
For model \eqref{eq-mixmodel1}, the   prior mean is given by
\begin{align}
E(f(X))&=\int  g(X,  G, \boldsymbol\kappa) E(P(\dif \bkappa \dif G)) =\int  g(X,  G, \boldsymbol\kappa) P_0(\dif \bkappa \dif G),
\intertext{while for model \eqref{eq-mixmodel2}, this is}
E(f(X))&=\int  g(X,  G, \boldsymbol\kappa)\mu(\dif \bkappa) P_0(\dif G).
\end{align}
The parameter $\alpha$ controls the concentration of the prior around the mean, and one can place a hyperprior on this as well.

In the following, we set $g(X,G,\bkappa)$ to be the matrix Langevin distribution with parameter $F = G\bkappa$. Thus,
\begin{equation}
g(X,G,\boldsymbol\kappa)=\etr(\bkappa G^TX)/Z(\bkappa)=C(\bkappa)\etr(\bkappa G^TX),
\end{equation}
with $C(\bkappa)=1/Z(\bkappa)=1/ _{0}F_1(\frac{1}{2}d, \frac{1}{4}\bkappa^T\bkappa)$.
Note that we have restricted ourselves to the special case where the matrix Langevin parameter $F$ has orthogonal columns (or equivalently, where $H = I_p$). While it
 is easy to apply our ideas to the general case, we demonstrate below that even with this restricted kernel, our nonparametric model has properties like
large support and consistency.

\subsection{Posterior consistency}
With our choice of parametric kernel, a DP prior on $\Pi_1$ induces
an infinite mixture of matrix Langevin distributions on $\mathcal{M}$. Call this distribution $\Pi$;
below, we show that this has large support on $\mathcal{M}$, and
that the resulting posterior distribution concentrates around  any true data generating density in $\mathcal{M}$. Our modelling framework and theory builds on \cite{abs1,abs2}, who developed consistency theorems for density estimation on compact Riemannian manifolds, and considered DP mixtures of kernels appropriate to the manifold under consideration.  However, they only considered simple manifolds, and showing that our proposed models have large support and consistency properties requires substantial new theory.

We first introduce some notions of distance and  neighborhoods on  $\mathcal{M}$.
A weak neighborhood  of $f_0$ with radius $\epsilon$ is defined as
\begin{equation}
\label{eq-weaknb}
W_{\epsilon}(f_0)=\left\{f: \left|\int zf\lambda(\dif X)- zf_0\lambda(\dif X)\right|\leq \epsilon, \text{for all}\; z\in C_b(V_{p,d}) \right\},
\end{equation}
where $C_b(V_{p,d})$ is the space of all continuous and bounded functions on $V_{p,d}$.
The  Hellinger  distance $d_H(f,f_0)$ is defined as
\begin{eqnarray*}
d_H(f,f_0) = \left(\dfrac{1}{2}\int (\sqrt{f(X)}-\sqrt{f_0(X)})^2\lambda(\dif X)\right)^{1/2}.
\end{eqnarray*}
 We let $U_{\epsilon}(f_0)$ denote  an $\epsilon$-Hellinger neighborhood around
$f_0$ with respect to $d_H$.  The Kullback-Leibler (KL)  divergence between $f_0$ and $f$  is defined to be
\begin{align}
\label{eq-KLdivergence}
d_{KL}(f_0,f)=\int  f_0(X) \log \dfrac{ f_0(X)}{f(X)}\lambda(\dif X),
\end{align}
with $K_{\epsilon}(f_0)$ denoting an $\epsilon$-KL neighborhood of $f_0$.

Let $X_1,\ldots, X_n$ be $n$ observations drawn i.i.d.\ from some true density $f_0$ on $V_{p,d}$.
 Under our model, the posterior probability $\Pi_{n}$ of some neighborhood  $W_\epsilon(f_0)$ is given by
\begin{align}
\label{eq-posteq}
\Pi_{n}\left(W_\epsilon(f_0)|X_1,\ldots, X_n\right)&=\dfrac{\int_{W_\epsilon(f_0)} \prod_{i=1}^n f(X_i)\Pi(\dif f)}{\int_{\mathcal{M}} \prod_{i=1}^n f(X_i)\Pi(\dif f)}.
\end{align}
 The posterior is  weakly consistent if for all $\epsilon>0$,  the following holds:
\begin{equation}
\Pi_{n}\left(W_\epsilon(f_0)|X_1,\ldots, X_n\right)\rightarrow 1 \;a.s. \;Pf_0^{\infty}\; \text{as}\; n\rightarrow \infty,
\end{equation}
where $Pf_0^{\infty}$ represents the true probability measure for $(X_1, X_2,\ldots)$.


We assume the  true density $f_0$ is continuous with $F_0$  as its probability distribution. The following theorem is on the weak consistency of the posterior under the mixture prior for both models \eqref{eq-mixmodel1} and  \eqref{eq-mixmodel2}, the proof of which is included in the appendix.
\begin{theorem}
\label{th-weakConsistency}
The posterior $\Pi_{n}$ in the DP-mixture of matrix Langevin distributions is weakly consistent. 
\end{theorem}

We now consider the consistency property of the posterior $\Pi_n$ with respect to the Hellinger neighborhood $U_{\epsilon}(f_0)$; this is referred as strong consistency.
%
\begin{theorem}
\label{th2}
Let $\pi_{\boldsymbol{\kappa}}$ be the prior on $\boldsymbol\kappa$, and
let $\Pi$ be the prior on $\mathcal{M}$ induced by $\Pi_1$ and $\pi_{\boldsymbol{\kappa}}$ via the mixture model \eqref{eq-mixmodel2}.
Let $\Pi_1\sim DP_{\alpha P_0}$ with $P_0$ a base measure having full support on $V_{p,d}$.  Assume $\pi_{\boldsymbol{\kappa}}(\phi^{-1}(n^a,\infty))\leq \exp(-n\beta)$ for some $a<1/((p+2)dp)$ and $\beta>0$ with $\phi(\boldsymbol{\kappa})=\sqrt{\sum_{i=1}^p(\kappa_i+1)^2}$. Then the posterior $\Pi_{n}$ is consistent with respect to the Hellinger distance $d_H$.
\end{theorem}

\begin{remark}
For prior $\pi_{\boldsymbol{\kappa}}$ on the concentration parameter $\boldsymbol{\kappa}$, to satisfy the condition
$\pi_{\boldsymbol{\kappa}}\left( \phi^{-1}(n^a,\infty)  \right)<\exp(-n\beta)$,
for some $a<1/(dp(p+2))$ and $\beta>0$ requires fast decay of the tails for $\pi_{\boldsymbol{\kappa}}$. One can check that an independent Weibull prior for $\kappa_i$, $i=1,\ldots, p$ with
$\kappa_i\sim \kappa_i^{\left(1/a\right)-1}\exp(-b \kappa_i^{(1/a)})$ will satisfy the tail  condition.

Another choice is to allow $\pi_{\boldsymbol{\kappa}}$ to be sample size dependent as suggested by \cite{abs2}. In this case, one can choose independent Gamma priors for
$\kappa_i$ with $\kappa_i\sim\kappa_i^{c}\exp(-b_n\kappa_i)$ where $c>0$ and $n^{1-a}/b_n\rightarrow 0$ with $0<a<1/(dp(p+2)).$
\end{remark}

\section{Inference for the nonparametric model}
\label{sec:post_sim}

A common approach to posterior inference for the Dirichlet process is Markov chain Monte Carlo
based on the Chinese restaurant process (CRP) representation 
of the DP \citep{Nea2000}.
The Chinese restaurant process describes the distribution over partitions of observations that results from integrating out the 
random probability measure $\Pi_1$, and a CRP-based Gibbs sampler updates this partition by reassigning each observation to a cluster conditioned on the
rest. The probability of an observation $X_i$ joining a cluster with parameters $(G,\bkappa$) is proportional to
the likelihood $g(X_i,g,\bkappa)$ times the number of observations already in that cluster (for an empty cluster, the latter is the
concentration parameter $\alpha$). Our case is complicated by the intractable likelihood $g(\cdot)$; this also makes updating
the cluster parameters  not straightforward. One possibility is to use an asymptotic approximation to the normalization constant $Z(\bkappa$) \citep{hoff2009}.
We instead use a recently proposed data augmentation scheme by \cite{rao14} to construct a Markov chain with the exact stationary distribution, and refer
the reader to that paper for details.

Below, we apply our nonparametric model to a dataset of near-Earth astronomical objects (comets and asteroids). 
Inferences were based on $5,000$ samples from the MCMC sampler, after a burn-in period of $1,000$ samples.

\subsection{Near Earth Objects dataset}

\begin{figure}
  \centering
    \includegraphics[width=.4\textwidth]{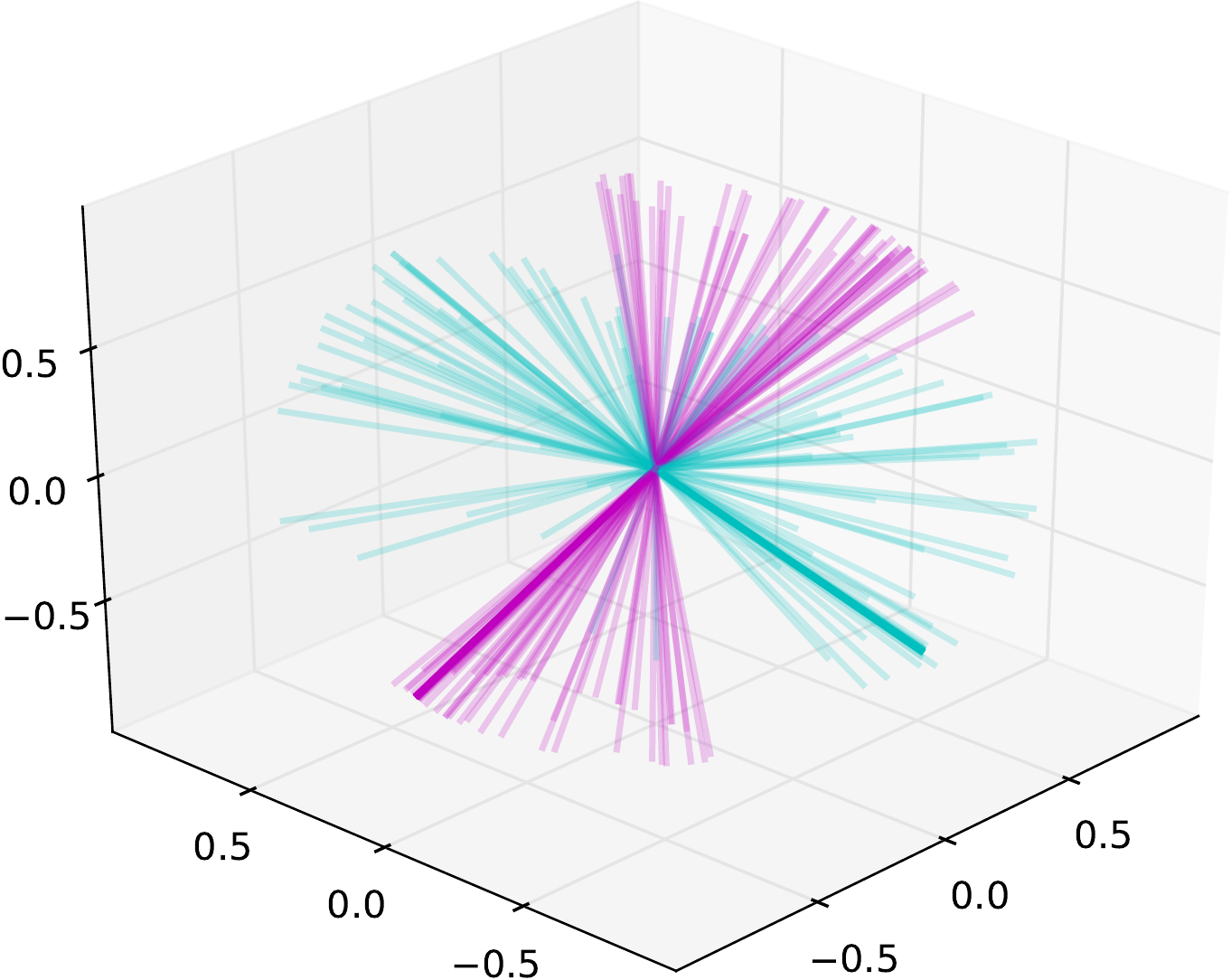}
  \centering
    \includegraphics[width=.3\textwidth]{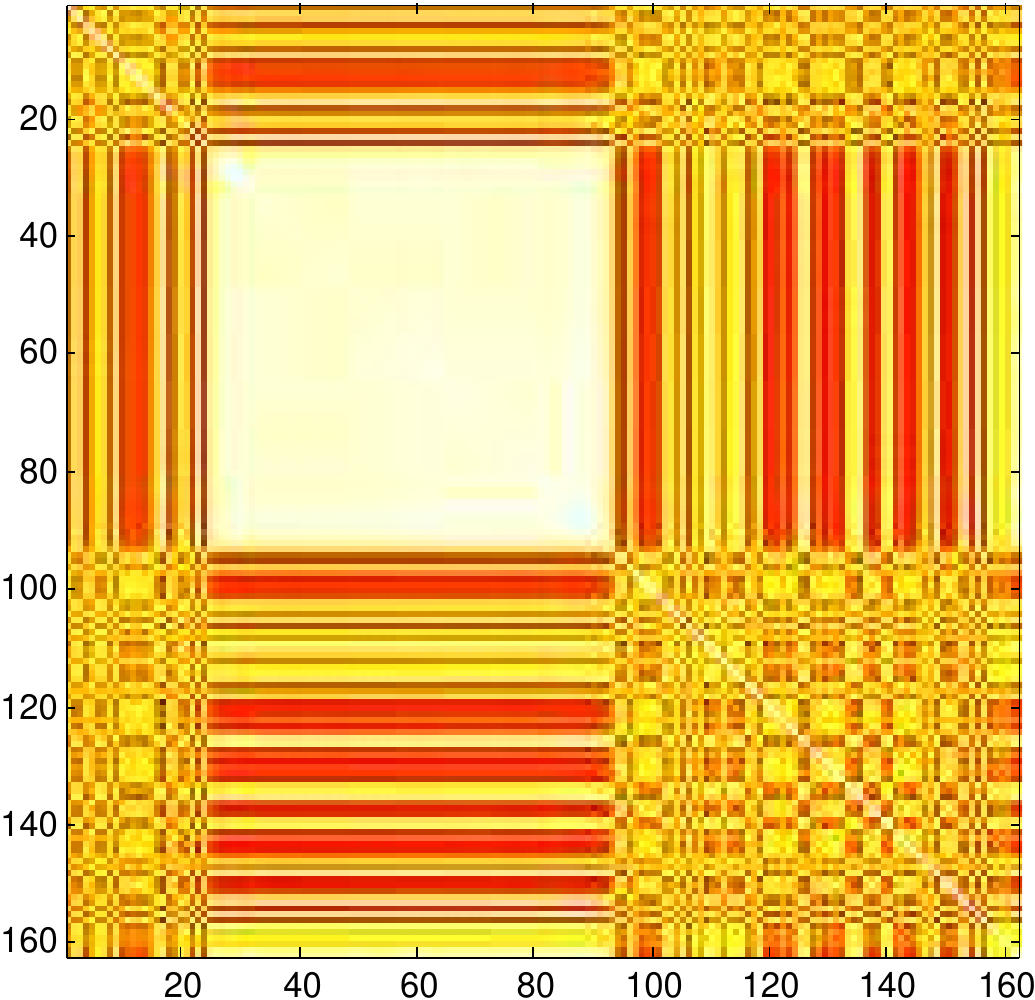}
\caption{The Near Earth Objects dataset (left), and the adjacency matrix inferred by the DP mixture model (right)}
  \label{fig:neo_data}
  \end{figure}

The Near Earth Objects dataset was collected by the
Near Earth Object Program of the National Aeronautics and Space Administration\footnote{Downloaded from
$\mathtt{http://neo.jpl.nasa.gov/cgi-bin/neo\_elem}$}, and consists of $162$ observations. Each data point
lies on the Stiefel manifold $V_{3,2}$, and characterizes the orientation of a two-dimensional elliptical orbit in
three-dimensional space.
The left subplot in Figure \ref{fig:neo_data} shows these data, with each $2$-frame represented as two orthonormal unit vectors.
The first component (representing the latitude of perihelion) is the set of cyan lines arranged as two horizontal cones.
The magenta lines (arranged as two vertical cones) form the second component, the longitude of perihelion.
  \begin{figure}
  \centering
    \includegraphics[width=.35\textwidth]{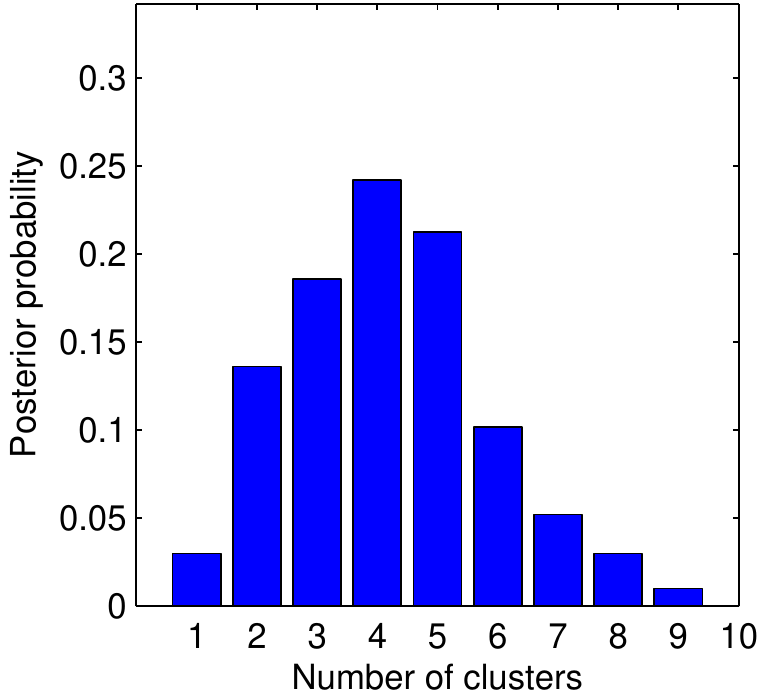}
  \centering
    \includegraphics[width=.42\textwidth]{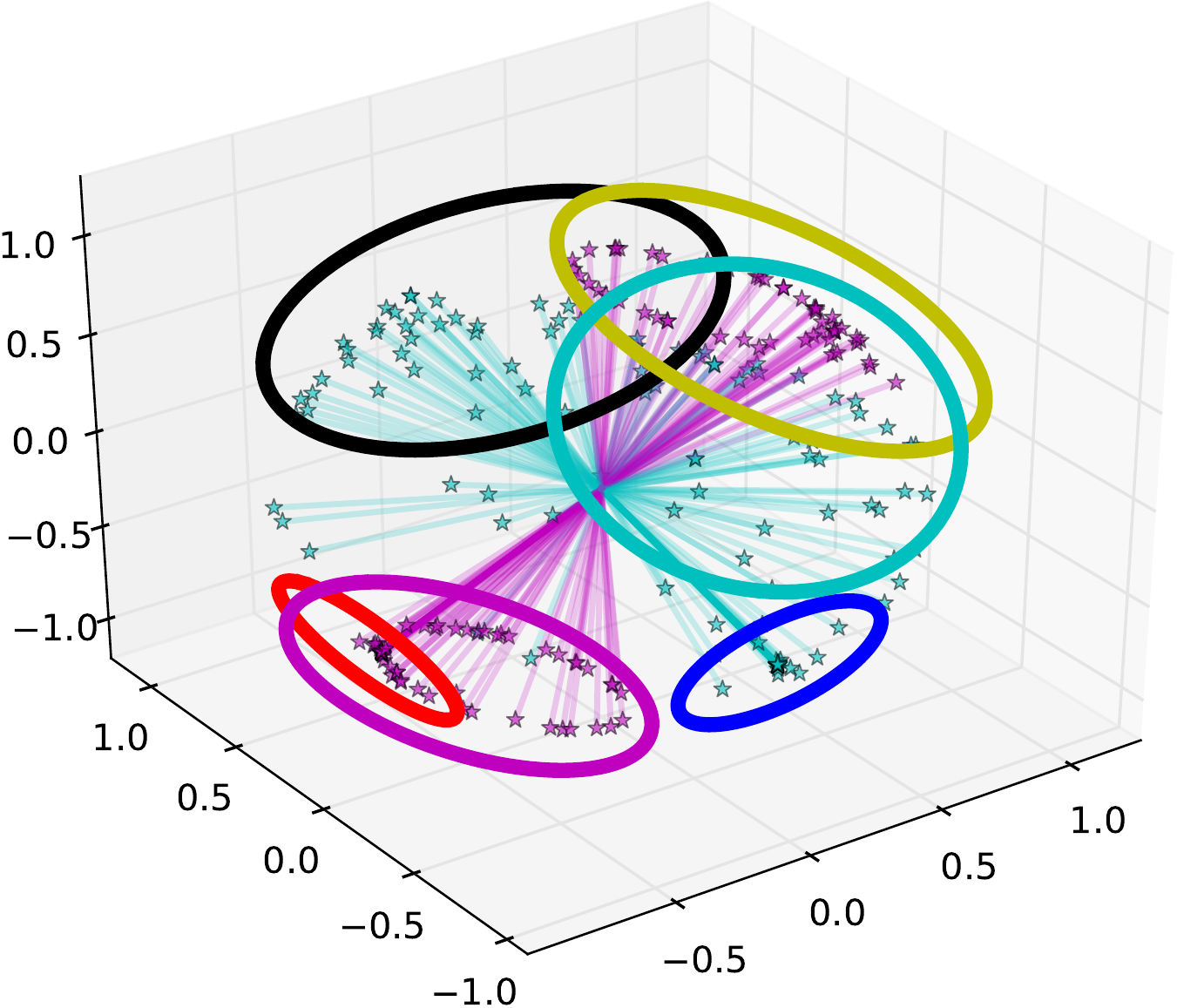}
\caption{Posterior over the number of clusters for the Near Earth Objects dataset (left), and location and scale parameters of an MCMC sample with
three clusters (right). The circles associated with each cluster correspond to $75\%$ predictive probability regions for the associated component.}
  \label{fig:post_nc}
  \end{figure}

We model this dataset as a DP mixture of matrix Langevin distributions. We set the DP concentration parameter $\alpha$
to $1$, and
for the DP base measure, 
placed independent probability measures on the matrices $G$ and $\bkappa$. 
For the former, we used a uniform prior (as in Section \ref{sec:Bays_inf}); 
however we found that an uninformative prior on $\bkappa$ resulted in high posterior probability for a single diffuse cluster
with no interesting structure.
To discourage this, we sought to penalize small values of $\kappa_i$. One way to do this is to
use a Gamma prior 
with a large shape parameter. Another is to use a hard constraint to bound the
$\kappa_i$'s away from small values. We took the latter approach, placing independent exponential priors restricted to $[5,\infty)$ on the
diagonal elements of $\bkappa$.

The right plot in Figure \ref{fig:neo_data} shows 
the adjacency matrix summarizing
the posterior distribution over clusterings. An off-diagonal element $(i,j)$ gives the number of times observations $i$ and $j$ were assigned to the
same cluster under the posterior. 
We see a highly coupled set of observations (from around observation $20$ to $80$ keeping the ordering of the downloaded dataset). This cluster
corresponds to a tightly grouped set of observations, visible as a pair of bold lines in the left plot of Figure \ref{fig:neo_data}.

  \begin{figure}
  \centering
    \includegraphics[width=.48\textwidth]{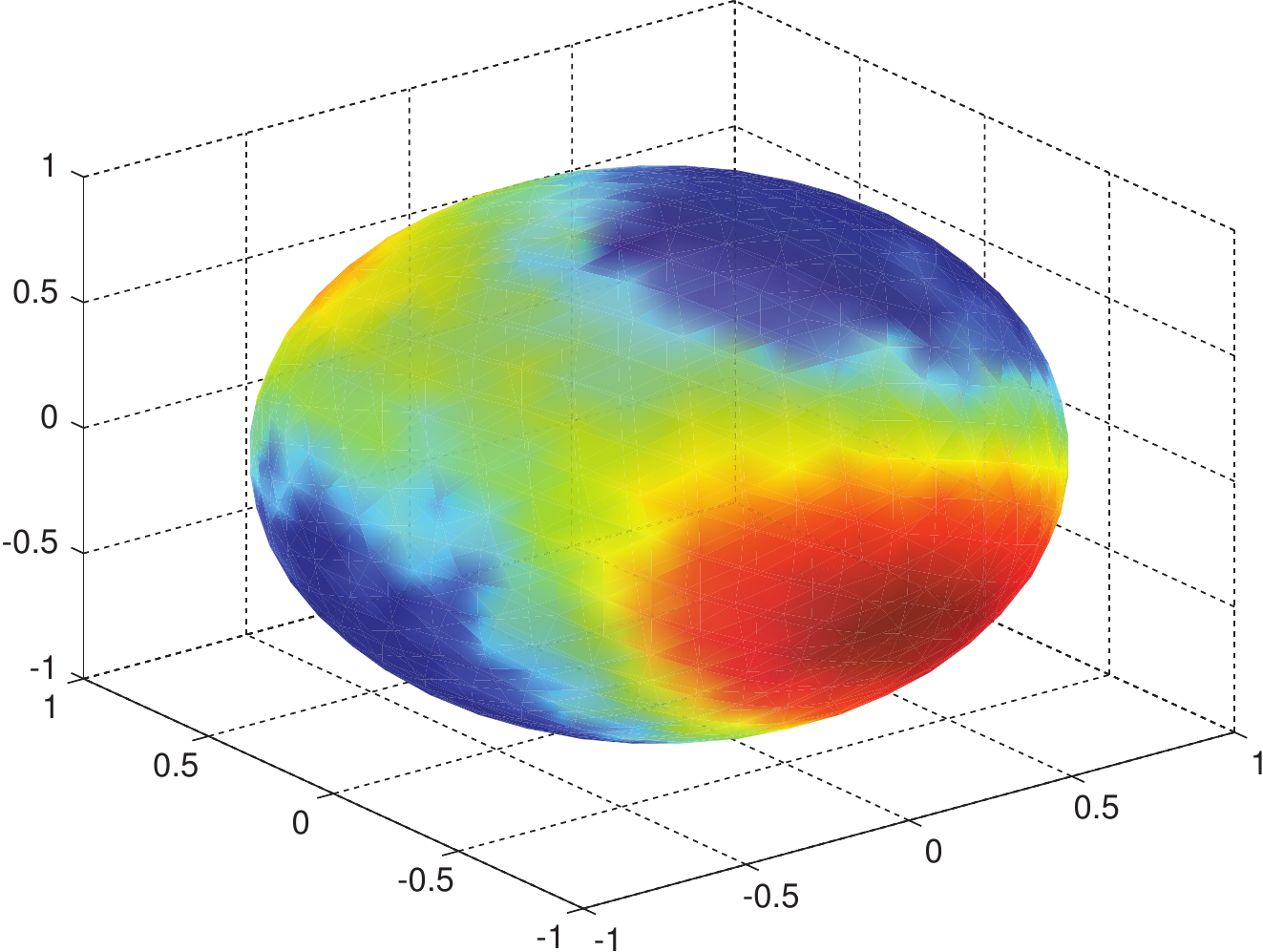}
  \centering
    \includegraphics[width=.48\textwidth]{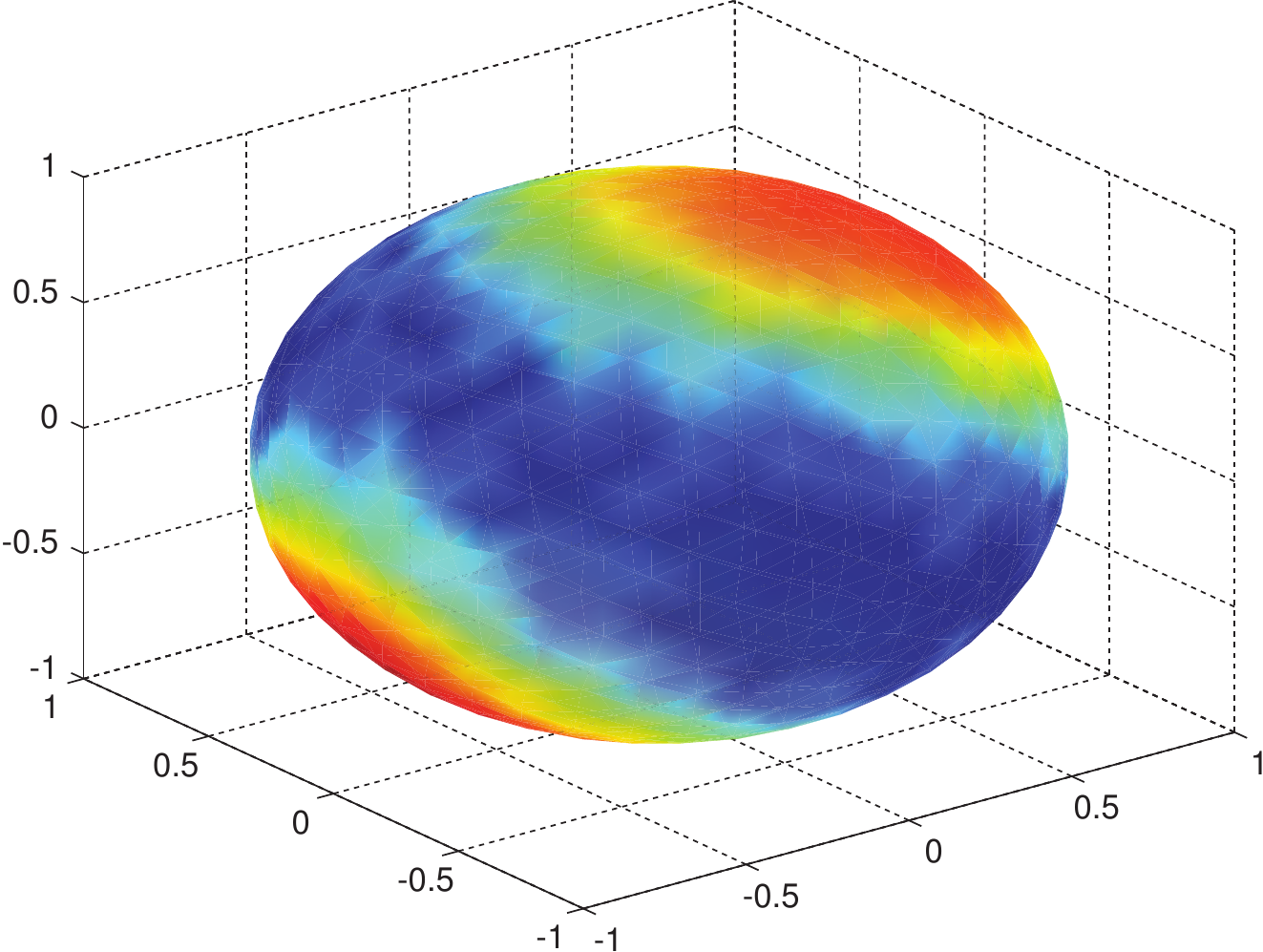}
\caption{Log predictive probabilities of first and second orthonormal components.}
  \label{fig:samplers_clst}
  \end{figure}
To investigate the underlying structure more carefully, we plot in Figure \ref{fig:post_nc} the posterior distribution over the number of
clusters. 
The figure shows this number is peaked at $4$, extending up to $9$. However, in most instances, most clusters
have a  small number of observations, with the posterior dominated by $2$ or $3$ large clusters.
A typical two cluster realization is fairly intuitive, with
each cluster corresponding to one of the two pairs of cones at right angles, and these
clusters were identified quite consistently across all posterior samples.
Occasionally, one or both of these might be further split into two smaller clusters, resulting in $3$ or $4$
clusters. A different example of a three cluster structure is shown in the right subfigure (this instance corresponded to the last MCMC sample of our chain
that had three large clusters).
In addition to the two aforementioned clusters,
this assigns the bunched group of observations mentioned earlier to their own cluster.
Parametric analysis of this dataset typically requires identifying this cluster and treating it as a single observation \citep{Sei2013440}; by contrast, our
nonparametric approach handles this much more naturally.

Finally, Figure \ref{fig:samplers_clst} show the log predictive-probabilities of observations given this dataset, with the left subplot giving the distribution of
the first component, and the right, the second. The peak of this distribution (the red spot to the right for the first plot, and the spot to the bottom left for the second),
correspond to the bunched set of observations mentioned earlier.
%


\section{Appendix}
In this appendix, we include the proofs for Theorem \ref{th-weakConsistency} and Theorem \ref{th2}.

Proof of Theorem \ref{th-weakConsistency}.
\begin{proof}
 The main ideas of proving consistency \citep[from][]{schwartz} are to bound the numerator of equation \eqref{eq-posteq} from above and the denominator from
below.  In order to bound the numerator, we construct uniformly consistent tests which separate the true density from its complement.
A condition on the prior mass of the Kullback-Leiber neighborhood of the true density is imposed to lower bound the denominator.  For weak consistency, it suffices to check  that the prior $\Pi$ assigns positive mass to any KL neighborhood of $f_0$ with which one can identify neighborhoods of $f_0$ for which uniformly consistent tests exist.

By slight abuse of notation, denote $f$ as any continuous function on $\cM$ in this proof.  From \cite{abs2} the following conditions are sufficient to verify that the KL support condition holds.  \begin{itemize}
\item [(1)] The kernel $g(X,G,\boldsymbol{\kappa}) $ is continuous in all of its arguments.
\item [(2)] The set $\{F_0\}\times D_{\epsilon}^{o}$ intersects the support of $\Pi_1 \times \pi_{\bkappa}$ with $D_{\epsilon}^{o}$ as the interior of $D_{\epsilon}$,
which is a compact neighborhood of some $\{{\kappa}_1,\ldots, {\kappa}_p  \}$ in $\mathbb {R}^p$.
\item [(3)] For any continuous function $f$ on $M$, there exists a compact neighborhood $D_{\epsilon}$ of $\{{\kappa}_1,\ldots, {\kappa}_p  \}$, such that
\begin{align*}
\sup_{X\in V_{p,d},\;\boldsymbol\kappa\in D_{\epsilon} } \bigg\|f(X)-\int g(X, G, \boldsymbol \kappa)f(G)\lambda (dG)\bigg\|\leq \epsilon.
\end{align*}
\end{itemize}

We first verify condition (1).
Note that one can write
\begin{align*}
g(X, G, \boldsymbol \kappa)&=C(\boldsymbol{\kappa})\etr(F^TX)= C(\boldsymbol{\kappa})\exp\left(\sum_{i=1}^p\kappa_i G_{[:i]}^TX_{[:i]}  \right).
\end{align*}
$g$ is continuous with respect to $\boldsymbol \kappa$ since the hypergeometric function $C(\boldsymbol \kappa)$ is continuous and  $\etr(F^TX)$ is clearly continuous with respect to $\boldsymbol{\kappa}$ as the exponential term can be viewed as a linear combination of $\kappa_i$'s.

Now rewrite the density as
\begin{align*}
g(X, G, \boldsymbol \kappa)=C(\boldsymbol{\kappa})\etr(F^TX)=C(\boldsymbol{\kappa})\exp\left(\dfrac{p+\sum_{i=1}^p \kappa_i^2-\rho(F, X)^2}{2} \right),
\end{align*}
where $\rho$ is the Frobenius distance between two matrices $F$ and $X$. Therefore $\etr(F^TX)$ is a continuous density of $X$ with respect to the Frobenius distance. As mentioned in section 2, $V_{p,d}$ can be embedded  onto the Euclidean space $M(d,p)$ via the inclusion map. Therefore, one can equip $V_{p, d}$ with a metric space structure via the extrinsic distance $\rho$ in the Euclidean space.
From the symmetry between $G$ and $X$, $g$ is also continuous with respect to $G$.

To prove (2), note that DP  has weak support on all the measures  whose support is contained by the base measure $P_0$
(See Theorem 3.2.4 in \cite{jk}, pp. 104). As $P_0$ and $\pi_{\bkappa}$ have full support, (2) follows immediately.

Let $I(X)=f(X)-\int g(X, G, \boldsymbol \kappa)f(G)\lambda (dG)$.  For the last condition, we must show that there exists
some compact subset in $\mathbb R^p$ with non-empty interior, $D_{\epsilon}$, such that
\begin{equation}
\sup_{X\in V_{p,d},\;\boldsymbol\kappa\in D_{\epsilon} } \|I(X)\|\leq \epsilon.
\end{equation}


From symmetry of $g$ with respect to $G$ and $X$, one can rewrite
\begin{align*}
I(X)&=C(\boldsymbol \kappa)\int \left(f(X)-f(G)\right)\etr(F^TX)\lambda(dG).
\end{align*}
Let $\widehat{G}=Q(d)^TG$ where $Q(d)$ is an orthogonal matrix with first $p$ columns being $X$. Then $G=Q(d)\widehat{G}$. As the volume form is invariant under the group action of the orthogonal matrices $O(d)$ on the left, then one has $\lambda(dG)=\lambda(d\widehat{G})$.
First note that
\begin{align*}
\rho^2(X, Q(d)\widehat{G})&=\Tr\left( \left(X-  Q(d)\widehat{G}\right)\left(X- Q(d)\widehat{G}\right)^T \right)\\
&=2p-2\Tr\left(  Q(d)^TX\widehat{G}^T \right)\\
&=2 \sum_{i=1}^p \left(1-\widehat{g}_{ii} \right),
\end{align*}
with $\widehat{g}_{ii}$ being the diagonal elements of $\widehat{G}$.
Let $(1-\widehat{g}_{ii})=\frac{1}{\kappa_i}s_{ii}$ for $i=1,\ldots,p$, with $s_{ii}\in [0, 2\kappa_i]$.
Then $\rho^2(X, Q(d)\widehat{G})=2\sum_{i=1}^p \frac{1}{\kappa_i}s_{ii}.$
As $\kappa_i\rightarrow \infty$ for all $i=1,\ldots, p$,  $\rho^2(X, Q(d)\widehat{G})\rightarrow 0$.
Since $f$ is continuous and the Stiefel manifold is compact, one has for any $\boldsymbol{s}=\{s_{11},\ldots, s_{pp}\}$,
\begin{align}
\label{eq-c1}
\sup_{X\in V_{p,d}} \left|\left(f(X)-f(Q(d)\widehat{G})\right)\right|\rightarrow 0,
\end{align}
as $\kappa_i\rightarrow \infty$ for all $i=1,\ldots, p$.
Let $\widehat{F}$ be the matrix whose $k$th column is $\kappa_kQ(d)\widehat{G}_{[:k]}.$  One has
\begin{align}
\label{eq-supterm}
\sup_{X\in V_{p,d}} |I(X)|&\nonumber\leq \sup_{X\in V_{p,d}}C(\boldsymbol \kappa)\int \left|\left(f(X)-f(Q(d)\widehat{G})\right)\right|\etr(\widehat{F}^TX)\lambda(d\widehat{G})\\ \nonumber
&\leq C(\boldsymbol \kappa)\int \left\{\sup_{X\in V_{p,d}} \left|\left(f(X)-f(Q(d)\widehat{G})\right)\right|\right\}\exp\left(\sum_{i=1}^p\kappa_i\widehat{g}_{ii} \right)\lambda(d\widehat{G})\\
&=C(\boldsymbol \kappa)\exp\left(\sum_{i=1}^p\kappa_i\right)\int \left\{\sup_{X\in V_{p,d}} \left|\left(f(X)-f(Q(d)\widehat{G})\right)\right|\right\}\exp\left(-\sum_{i=1}^p s_{ii} \right)\lambda(d\widehat{G}).
\end{align}
Let $\pi_1$ be the transformation given by $\pi_1(\widehat{g}_{ij})=\widehat{g}_{ij}$ when $i\neq j$ and $\pi_1(\widehat{g}_{ii})=s_{ii}=\kappa_i(1-\widehat{g}_{ii}).$  Denote $\lambda(d\widehat{G_s})$ as new volume measure after changing of variables with respect to $\pi_1$. Let $J_1$ be the Jacobian of the map $\pi_1$. Rewrite $\lambda(d\widehat{G})=\varphi(\widehat{G})d\widehat{g}_{11}\wedge d\widehat{g}_{12}\cdots\wedge\widehat{g}_{dp}$ where $\varphi(\widehat{G})$ is some function of $\widehat{G}$.  Then $\lambda(d\widehat{G})$ is given by the pullback of $\lambda(d\widehat{G}_s)$ induced by the map $\pi_1$, that is
\begin{align}
\lambda(d\widehat{G})=(\pi_1)^{*}(\lambda(d\widehat{G}_s))=\varphi(\pi_1(\widehat{G}))\frac{1}{\det(J_1)}ds_{11}\wedge ds_{12}\cdots \wedge ds_{dp},
\end{align}
where $s_{ij}$ is the $(i,j)$th element of $\widehat{G}_s$.

Then the last term of \eqref{eq-supterm} becomes
\begin{align}
&\nonumber C(\boldsymbol \kappa)\exp\left(\sum_{i=1}^p\kappa_i\right)\prod_{i=1}^p\dfrac{1}{\kappa_i}\int \left\{\sup_{X\in V_{p,d}} \left|\left(f(X)-f(Q(d)\widehat{G})\right)\right|\right\}\times\\
&\exp\left(\sum_{i=1}^p-s_{ii} \right)\frac{1}{\det(J_1)}\lambda(d\widehat{G_s}),
\end{align}
with appropriate change of the range of integration. 
It is not hard  to see that
\begin{equation}\int \exp\left(-\sum_{i=1}^p s_{ii} \right)\frac{1}{\det(J_1)}\lambda(d\widehat{G_s})<\infty.
\end{equation}

We now proceed to show that even as $\kappa_i \rightarrow \infty$,
\begin{equation}
C(\boldsymbol \kappa)\exp\left(\sum_{i=1}^p\kappa_i\right)\prod_{i=1}^p\dfrac{1}{\kappa_i}<\infty.
\end{equation}
One has
\begin{align*}
C(\boldsymbol \kappa)\exp\left(\sum_{i=1}^p\kappa_i\right)\prod_{i=1}^p\dfrac{1}{\kappa_i}
&=\dfrac{\prod_{i=1}^{p}\dfrac{1}{\kappa_i}}{_{0}F_1\left(\frac{1}{2}d, \frac{1}{4}\diag \left\{\kappa_1^2,\ldots, \kappa_p^2  \right\}\right)/\prod_{i=1}^p\exp(\kappa_i)}.
\end{align*}
Write (see \cite{bultler})
\begin{equation}
_{0}F_1\left(\frac{1}{2}d, \frac{1}{4}\diag \left\{\kappa_1^2,\ldots, \kappa_p^2  \right\}\right)=\int _{O_p}\etr\left(\diag \left\{\kappa_1,\ldots, \kappa_p  \right\}T  \right)dT
\end{equation}
with $T\in O_p$ the group of all the $p$ by $p$ orthogonal matrices with $dT$ given by $\wedge_{i<j}t_j^Tdt_i$.
When $\kappa_i\geq 1$ for $i=1,\ldots,p$, one looks at
\begin{align*}
\int _{O_p}\frac{\etr\left(\diag \left\{\kappa_1,\ldots, \kappa_p  \right\}T  \right)}{\prod_{i=1}^p\exp(\kappa_i)}dT&=\int_{O_p}\exp\left(-\left(\sum_{i=1}^p \kappa_i(1-t_{ii})  \right)\right)dT,
\end{align*}
where $t_{ii}$ are the diagonal elements of $T$.  Let $\pi_2$ be the of changing of variable with $u_{ii}=\kappa_i(1-t_{ii})$, one has $u_{ii}\in [0, 2\kappa_i]$. Let $d\widehat{T}$ be the volume form after changing of variable. We then have
\begin{align*}
\int_{O_p}\exp\left(-\left(\sum_{i=1}^p \kappa_i(1-t_{ii})  \right)\right)dT=\prod_{i=1}^p1/\kappa_i\int \exp\left(-\left(\sum_{i=1}^p u_{ii}  \right)\right)\frac{1}{\det(J_2)}d\widehat{T},
\end{align*}
where $\det(J_2)$ corresponding to determinants of the Jacobian of maps $\pi_2$  which is essentially the same map as $\pi_1$  but with domain $T\in O_p$.
Note $\int \exp\left(-\left(\sum_{i=1}^p u_{ii}  \right)\right)\frac{1}{\det(J_2)}d\widehat{T}$ is bounded away from zero and infinity as $\kappa_i\rightarrow \infty$. Therefore, we can conclude
\begin{equation}
\label{eq-c2}
C(\boldsymbol \kappa)\exp\left(\sum_{i=1}^p\kappa_i\right)\prod_{i=1}^p\dfrac{1}{\kappa_i}<\infty.
\end{equation}
Therefore, combining \eqref{eq-c1} and \eqref{eq-c2} and by the dominated convergence theorem, one has
\begin{align*}
\sup_{X\in V_{p,d}} |I(X)|\rightarrow 0
\end{align*}
as $\kappa_i\rightarrow \infty$ for all $i=1,\ldots, p$. Thus  for all $\epsilon>0$, there exists $M_i$ large enough such that, when $\kappa_i>M_i$, $\sup_{X\in V_{p,d}} |I(X)|\leq \epsilon$. One can take $D_{\epsilon}$ to be a $\epsilon$ neighborhood of $\{\kappa_1,\ldots, \kappa_p\}$ with $\kappa_i>\max\{M_i,i=1,\ldots, p\}$.
\end{proof}


Proof of Theorem \ref{th2}.

\begin{proof}
In order to establish strong consistency, it is not sufficient for the prior $\Pi$ to assign positive mass to any Kullback-Leibler neighborhood of $f_0$. 
We need to construct high mass sieves with   metric entropy $N(\epsilon, \mathcal{F})$ bounded by certain order where $N(\epsilon, \mathcal{F})$  is defined as the logarithm of the
minimum number of balls with Hellinger radius $\epsilon$ to cover the space $\mathcal{F}$.  We refer to \cite{barron} for some general strong consistency theorems.
We first proceed to verify the following two conditions on the kernel $g(X,G, \boldsymbol{\kappa})$.
\begin{itemize}
\item [(a)] There exists positive constants $k_0$, $a_1$ and $A_1$ such that for all $k>k_0$, $G_1,G_2\in V_{p,d}$ one has
\begin{equation}
\label{eq-ca}
\sup_{X\in V_{p,d}, \boldsymbol{\kappa}\in \phi^{-1}[0,k]} |g(X,G_1,\boldsymbol{\kappa})-g(X,G_2,\boldsymbol{\kappa})|\leq A_1k^{a_1}\rho (G_1,G_2),
\end{equation}
where $\phi: \mathbb R^{p}\rightarrow [0,\infty)$ is some  continuous function of $\boldsymbol{\kappa}$.
\item [(b)] There exists positive constants $a_2$ and $A_2$ such that for all $\boldsymbol{\kappa}, \widetilde{\boldsymbol{\kappa}} \in \phi^{-1}[0,k]$, $k\geq k_0$,
\begin{equation}
\label{eq-cb}
\sup_{X, G\in V_{p,d}} |g(X,G,\boldsymbol{\kappa})-g(X,G,\widetilde{\boldsymbol{\kappa}})|\leq A_2k^{a_2}\rho_2 (\boldsymbol{\kappa},\widetilde{\boldsymbol{\kappa}}),
\end{equation}
where $\rho_2$ is the Euclidean distance $\|\cdot\|_2$ on $\mathbb{R}^p$.
\end{itemize}

Let $G_1, G_2\in V_{p,d}$ and $F_1$ and $F_2$ be such that their $i$th columns are given by $\kappa_iG_{1_{[:,i]}}$ and $\kappa_iG_{2_{[:,i]}}$ respectively.
For $s, t\in [0,c] $  and $c>0$, one has
\begin{align*}
\Big|\exp\left(-\frac{s^2}{2}\right)-\exp\left(-\frac{t^2}{2}\right)\Big|\leq \Big|\eta\exp\left(-\frac{\eta^2}{2}\right)(s-t)\Big|\leq c |s-t|,
\end{align*}
where $\eta$ is some point between $s$ and $t$.
 Let $k_{\max}=\max\{\kappa_1,\ldots, \kappa_p\}$. A little calculation shows that $\rho(F,X)\leq \sqrt{\sum_{i=1}^p (\kappa_i+1)^2}$, so that
\begin{align*}
&\sup_{X\in V_{p,d}, \boldsymbol{\kappa}\in \phi^{-1}[0,k]} \bigg|g(X,G_1,\boldsymbol{\kappa})-g(X,G_2,\boldsymbol{\kappa})\bigg|\\
&=\sup_{X\in V_{p,d}, \boldsymbol{\kappa}\in \phi^{-1}[0,k]}\bigg|C(\boldsymbol{\kappa})\exp \left(\frac{p}{2}\right)\exp\left(\frac{\sum_{i=1}^{p}\kappa_i}{2}  \right)\left(\exp\left(-\frac{\rho^2(F_1,X)}{2}\right)- \exp\left(-\frac{\rho^2(F_2,X)}{2}\right) \right)\bigg|\\
&\leq \sup_{X\in V_{p,d}, \boldsymbol{\kappa}\in \phi^{-1}[0,k]}\bigg|C(\boldsymbol{\kappa})\exp \left(\frac{p}{2}\right)\exp\left(\frac{\sum_{i=1}^{p}\kappa_i}{2}  \right)\sqrt{\sum_{i=1}^p (\kappa_i+1)^2}\left(\rho(F_1,X)-\rho(F_2,X)\right)\bigg|\\
&\leq \exp \left(\frac{p}{2}\right)\sup_{X\in V_{p,d}, \boldsymbol{\kappa}\in \phi^{-1}[0,k]}\bigg|C(\boldsymbol{\kappa})\exp\left(\frac{\sum_{i=1}^{p}\kappa_i}{2}  \right)\rho(F_1,F_2)\sqrt{\sum_{i=1}^p (\kappa_i+1)^2}\bigg|\\
&\leq 2\exp \left(\frac{p}{2}\right)\sup_{X\in V_{p,d}, \boldsymbol{\kappa}\in \phi^{-1}[0,k]}\bigg|C(\boldsymbol{\kappa})\exp\left(\frac{\sum_{i=1}^{p}\kappa_i}{2}  \right)\sqrt{\sum_{i=1}^p\kappa_i^2}\rho(G_1,G_2)\sqrt{\sum_{i=1}^p (\kappa_i+1)^2}\bigg|\\
&\leq 2\exp \left(\frac{p}{2}\right)\sup_{X\in V_{p,d}, \boldsymbol{\kappa}\in \phi^{-1}[0,k]}\bigg| C\prod_{i=1}^p\kappa_i\sqrt{\sum_{i=1}^p\kappa_i^2}\sqrt{\sum_{i=1}^p (\kappa_i+1)^2} \rho(G_1,G_2)\bigg|
\end{align*}
where $C$ is some constant according to \eqref{eq-c2}.
Let $\phi(\boldsymbol{\kappa})=\sqrt{\sum_{i=1}^p (\kappa_i+1)^2}$. If $\phi(\boldsymbol{\kappa})\leq k$, then
$\sqrt{\sum_{i=1}^p\kappa_i^2}\leq \phi(\boldsymbol{\kappa})\leq k$ and $\kappa_i\leq k$ for each $i$. Thus $\prod_{i=1}^p\kappa_i\leq k^p$. Therefore,
\begin{align*}
\sup_{X\in V_{p,d}, \boldsymbol{\kappa}\in \phi^{-1}[0,k]} |g(X,G_1,\boldsymbol{\kappa})-g(X,G_2,\boldsymbol{\kappa})|\leq C_1k^{p+2}\rho(G_1,G_2),
\end{align*}
with $C_1$ some constant.
Let $a_1=p+2$, then condition (a) holds.

Let $\boldsymbol{\kappa}$, $\widetilde{\boldsymbol{\kappa}}\in \mathbb R^p$ be two vectors of the concentration parameters.
By the mean value theorem, one has for some $t\in (0,1)$
\begin{align*}
g(X,G,\boldsymbol{\kappa})-g(X,G,\widetilde{\boldsymbol{\kappa}})=\left(\bigtriangledown g(X,G, (1-t)\boldsymbol{\kappa}+t\widetilde{\boldsymbol{\kappa}})\right)\cdot(\boldsymbol{\kappa}-\widetilde{\boldsymbol{\kappa}}),
\end{align*}
where $\bigtriangledown g(X,G, (1-t)\boldsymbol\kappa+t\widetilde{\boldsymbol{\kappa}})$ is the gradient of $g(X,G, \boldsymbol\kappa)$ with respect to $\boldsymbol{\kappa}$ evaluated at $(1-t)\boldsymbol\kappa+t\widetilde{\boldsymbol{\kappa}}$  and $\cdot$ denotes the inner product.
By Cauchy-Schwarz inequality, one has
\begin{align*}
 |g(X,G,\boldsymbol{\kappa})-g(X,G,\widetilde{\boldsymbol{\kappa}})|\leq \| \bigtriangledown g(X,G, (1-t)\boldsymbol\kappa+t\widetilde{\boldsymbol{\kappa}}) \|_2  \| \boldsymbol{\kappa}-\widetilde{\boldsymbol{\kappa}}  \|_2.
\end{align*}
Note that  for $i=1,\ldots, p$,
\begin{align*}
\dfrac{\partial g}{\partial \kappa_i}&=\exp\left(-\sum_{i=1}^p \kappa_i(1-G_{[:i]}^TX_{[:i]} )\right) \left(C(\boldsymbol{\kappa})G_{[:i]}^TX_{[:i]}\exp(\sum_{i=1}^p\kappa_i)+\dfrac{\partial C(\boldsymbol{\kappa})}{\partial \kappa_i} \exp(\sum_{i=1}^p\kappa_i) \right)\\
&=\exp\left(-\sum_{i=1}^p \kappa_i(1-G_{[:i]}^TX_{[:i]} )\right) \bigg(C(\boldsymbol{\kappa})G_{[:i]}^TX_{[:i]}\exp(\sum_{i=1}^p\kappa_i)\\
&-C^2(\boldsymbol{\kappa})\dfrac{\partial _{0}F_1\left(\frac{1}{2}d, \frac{1}{4}\diag \left\{\kappa_1^2,\ldots, \kappa_p^2  \right\}\right)}{\partial \kappa_i}\exp(\sum_{i=1}^p\kappa_i)\bigg).
\end{align*}

By applying the general Leibniz rule for differentiation under an integral sign, one has
\begin{align*}
\dfrac{\partial _{0}F_1\left(\frac{1}{2}d, \frac{1}{4}\diag \left\{\kappa_1^2,\ldots, \kappa_p^2  \right\}\right)}{\partial \kappa_i}&=\int _{O_p}\dfrac{\partial\etr\left(\diag \left\{\kappa_1,\ldots, \kappa_p  \right\}S  \right)}{\partial \kappa_i}dS\\
&=\int _{O_p}s_{ii} \exp \left(\sum_{i=1}^p\kappa_is_{ii}\right)dS\\
&\leq \int _{O_p} \exp \left(\sum_{i=1}^p\kappa_is_{ii}\right)dS=\dfrac{1}{C(\boldsymbol{\kappa})}.
\end{align*}
Then one has \begin{align*}
\left|\dfrac{\partial g(X,G,\boldsymbol{\kappa})}{\partial \kappa_i}\right|&\leq C(\boldsymbol{\kappa})\exp\left(\sum_{i=1}^p\kappa_i\right)+C^2(\boldsymbol{\kappa})\dfrac{\partial _{0}F_1\left(\frac{1}{2}d, \frac{1}{4}\diag \left\{\kappa_1^2,\ldots, \kappa_p^2  \right\}\right)}{\partial\kappa_i}\exp\left(\sum_{i=1}^p\kappa_i\right)\\
&\leq 2C(\boldsymbol{\kappa})\exp\left(\sum_{i=1}^p\kappa_i\right)\leq C_2\prod_{i=1}^p\kappa_i,
\end{align*}
for some constant $C_2$ by \eqref{eq-c2}.
Therefore,
\begin{align*}
 \| \bigtriangledown g(X,G, (1-t)\boldsymbol\kappa+t\widetilde{\boldsymbol{\kappa}}) \|_2\leq C_2k^{p}.
\end{align*}
Then one has
\begin{align*}
 |g(X,G,\boldsymbol{\kappa})-g(X,G,\widetilde{\boldsymbol{\kappa}})|\leq C_2k^p \| \boldsymbol{\kappa}-\widetilde{\boldsymbol{\kappa}}  \|_2.
\end{align*}
Letting $a_2=p$, condition (b) is verified.

We proceed to verify the two following entropy conditions:
\begin{itemize}
\item [(c)] For any $k\geq k_0$, the subset $\phi^{-1}[0, k]$ is compact and its $\epsilon$-covering number is bounded by $(k\epsilon^{-1})^{b_2}$ for some constant $b_2$ independent of $\boldsymbol{\kappa}$ and $\epsilon$.

\item [(d)] The $\epsilon$ covering number of the manifold $V_{p,d}$ is bounded by $A_3\epsilon^{-a_3}$ for any $\epsilon>0$.
\end{itemize}
It is easy to verify condition (c) as $\phi^{-1}([0,k])=\{\boldsymbol{\kappa}, \sum_{i=1}^p(\kappa_i+1)^2\leq k^2  \}$, which is a subset of a shifted Euclidean ball in $\mathbb R^p$ with radius $k$. With a direct argument using packing numbers \citep[see Section 4]{pollard},
one can obtain a  bound for the entropy of $\phi^{-1}[0,k]$ which is given by $\dfrac{3k^p}{\epsilon^p}$. Thus condition (c) holds with $b_2=p$.

Denote $N(\epsilon)$ as the entropy of $V_{p,d}$ and $N_{E}(\epsilon)$ as the entropy of $V_{p,d}$ viewed as a subset of $\mathbb R^{pd}$
(thus points covering $V_{p,d}$ do not necessarily lie on $V_{p,d}$ for the latter case). One can show that $N(2\epsilon)\leq N_{E}(\epsilon)$.
Note that $V_{p,d}\subset[-1,1]^{pd}$ which is a subset of a Euclidean ball of radius $\sqrt{dp}$ centered at zero, the $\epsilon$ number of which is bounded $\left(\dfrac{3\sqrt{dp}}{\epsilon}\right)^{dp}$. Therefore, condition (d) holds with $a_3=dp$.  Then by Corollary 1 in \cite{abs2}, strong consistency follows.

\end{proof}

\bibliographystyle{apalike}
\bibliography{refs}

\end{document}